\documentclass{amsart}
\usepackage{fullpage,amssymb,epic,eepic,epsfig}
\usepackage{ipe}
\usepackage{latexsym}

\theoremstyle{plain}

\newtheorem{theorem}{Theorem}
\newtheorem{proposition}{Proposition}[section]

\theoremstyle{definition}
\newtheorem{definition}[proposition]{Definition}

\theoremstyle{remark}

\def\printname#1{
        \if\draft y
                \smash{\makebox[0pt]{\hspace{-0.5in}
                        \raisebox{8pt}{\tt\tiny #1}}}
        \fi
}

\newlength{\standardunitlength}
\catcode`\@=11
\long\def\@makecaption#1#2{%
     \vskip 10pt

\setbox\@tempboxa\hbox{
       \small\sf{\bfcaptionfont #1. }\ignorespaces #2}%
     \ifdim \wd\@tempboxa >\captionwidth {%
         \rightskip=\@captionmargin\leftskip=\@captionmargin
         \unhbox\@tempboxa\par}%
       \else
         \hbox to\hsize{\hfil\box\@tempboxa\hfil}%
     \fi}
\font\bfcaptionfont=cmssbx10 scaled \magstephalf
\newdimen\@captionmargin\@captionmargin=2\parindent
\newdimen\captionwidth\captionwidth=\hsize
\catcode`\@=12

\def\E{\mathcal E}

\def\sgn{\operatorname{sgn}}

\newcommand{\done}{\hfill$\Box$\par}

\def\ti{\widetilde}

\def\sign{\mathrm{sign}}

\begin{document}


\title[The 3D Ising Problem Revisited]
{The 3D Dimer and Ising Problems Revisited}

\author{Martin Loebl}
\address{Dept.~of Applied Mathematics and\\
Institute of Theoretical Computer Science (ITI)\\
Charles University \\
Malostransk\' e n. 25 \\
118 00 Praha 1 \\
Czech Republic.}
\email{loebl@kam.mff.cuni.cz}
\author{Lenka Zdeborov\' a}
\address{Institute of Physics \\
Academy of Science of the Czech Republic\\
Na Slovance 2\\
182 21 Praha 8\\
Czech Republic.}
\email{zdeborl@fzu.cz}

\thanks{The second author gratefully acknowledges the support of project AVOZ10100520 
of the Academy of Sciences of the Czech Republic. 
2000 {\em Mathematics Classification.} Primary 82B20. Secondary 82B41.
}

\date{
This edition: \today \hspace{0.5cm} First edition: October 2004.}


\begin{abstract}
We express the finite 3D Dimer partition function as a linear combination of determinants 
of oriented adjacency matrices, and the finite 3D Ising partition sum as a linear 
combination of products over aperiodic closed walks. The methodology we use is embedding 
of cubic lattice on 2D surfaces of large genus. 
\end{abstract}

\maketitle



\section{Introduction}
\label{sec.int}

The motivation for this work has been a renewed interest of theoretical physicists 
in basic statistical physics models, the Ising and the Dimer models.
Let us mention two such recent promising works.
The first one is a new relation (duality) of topological strings and the Dimer model, see e.g. 
\cite{ORV}.
The second one is a relation of the theory of discrete Riemann surfaces \cite{M}
and Theta functions with criticality of the Ising problem \cite{AMV, CM1, CM2}.

In this paper we present new formulas for the finite 3D Dimer and 3D Ising problems.
The formulas are obtained in a combinatorial way. We hope that a profitable relation
with the work mentioned above may be found.

A model we consider here is the
{\it Ising version of the Edwards-Anderson model}.
 It can be described as follows.
 A {\it coupling constant} $J_{ij}$ is assigned to each bond $\{i, j\}$
 of a given lattice graph G;
the coupling constant characterizes the interaction between the particles
represented by sites $i$ and $j$.
A physical state of the system is an assignment of spin
$\sigma _i \in \{+1,-1\}$ to each site $i$.
The {\it Hamiltonian} (or energy function) is defined as
$H(\sigma) = -\sum_{\{i,j\} \in E} {J_{ij} \sigma_i \sigma_j}$.
The distribution of physical states over all possible energy levels
is encapsulated in the {\it partition function}
$Z(\beta) = \sum_{\sigma} {e^{-\beta H(\sigma)}}$ from which
all fundamental physical quantities may be derived.

We may reformulate the Ising problem
in graph theoretic terms as follows.
 A graph is a pair $G=(V,E)$ where $V$ is a set of {\it vertices} and $E$ is
now the set of edges (not the energy).
 A graph with some regularity properties may be called a {\it lattice graph}.
 We associate with each edge $e$ of $G$
 a weight $w(e)$ and for a subset of edges $A \subset E$,
 $w(A)$ will denote the sum
of the weights $w(e)$ associated with the edges in $A$.

An {\it even subgraph} of a graph $G=(V,E)$ is a set of edges $U \subset E$
such that each vertex of $V$ is incident with an even number of edges from $U$.
The {\it generating function of even subgraphs} ${\mathcal H}(G,x)$
equals the sum of $\alpha^{w(U)}$ over all even subgraphs  $U$ of $G$. A classic
relation between the Ising partition function and the generating function of
even subgraphs of the same graph states that
$$
Z(\beta) = {2^n} \left( {\prod_{\{i,j\} \in E} \cosh({\beta} J_{ij})}
\right)  {\mathcal H} (G, \tanh(\beta J_{ij})).
$$

A subset of edges $P\subset E$ is called
a {\it perfect matching} or {\it dimer arrangement} if each vertex belongs
to exactly one element of $P$.
The {\it dimer partition function} on graph $G$ may be viewed as a polynomial
${\mathcal P}(G,\alpha)$ which equals the sum of $\alpha^{w(P)}$ over all perfect
matchings $P$ of $G$. This polynomial is also called the {\it generating function
of perfect matchings}.
\medskip

The generating functions of even subgraphs and perfect matchings may
be defined in
a more general way as follows: associate a variable $x_e$ with each edge
$e$ of graph $G$, let $x(A)=\prod_{e\in A} x_e$ and let e.g. the generating
function of perfect matchings be the sum of $x(P)$, $P$
perfect matching of $G$.
All results introduced in this paper also hold in this more general setting;
however the presentation using weights rather than variables is perhaps
more natural.
\medskip\noindent

\subsection*{Convention for the cubic lattice}
\label{sub.conv}

This paper studies properties of finite cubic lattices. Let us now fix
some notation for them.

\begin{figure}[ht]
\begin{minipage}{0.45\linewidth}
\includegraphics[width=6.5cm]{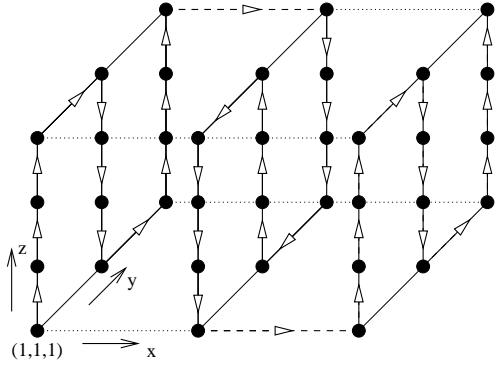}
\caption{\label{fig-orde} Illustration of the linear order of vertices
  on a lattice $Q(3,3,4)$.}
\end{minipage}
\begin{minipage}{0.50\linewidth}

Let $m_1, m_2, k$ be positive integers, where $k$ is even and $m_1$
and $m_2$ odd.
The cubic lattice $Q= Q(m_1m_2k)$ is the following graph: \\

\noindent The vertices are: \\
\noindent $V(xyz)$, $x= 1,\dots, m_1$, $y= 1,\dots, m_2$,
$z=1,\dots, k$. \\

\noindent The edges are:
\begin{itemize}
\item{the {\it vertical} edges $v_{xyz}=\{V_{xyz},V_{xy(z+1)}\}$,\\
    $z=1,\dots, k-1$,}
\item{the {\it width} edges  $w_{xyz}=\{V_{xyz},V_{x(y+1)z}\}$,\\ $y=1,\dots, m_2-1$, }
\item{the {\it horizontal} edges  $h_{xyz}=\{V_{xyz},V_{(x+1)yz}\}$,\\
$x=1,\dots, m_1-1$.}
\end{itemize}
\end{minipage}
\end{figure}

\begin{definition}
\label{lin_order}
We now fix as in Fig. \ref{fig-orde} a {\it linear order} '$<$' on
the set of the vertices of $Q$ as
$$
 V(11), {\bar V(12)},\dots, { V(1m_2)}, {\bar V(2m_2)}, {V(2(m_2-1))},
 \dots, { V(m_1 m_2)}
$$
where $V(xy)= V(xy1),\dots, V(xyk)$ and ${\bar V_{xy}}$
denotes the reversal of $V_{xy}$.
\end{definition}

\begin{figure}[ht]
\begin{minipage}{0.38\linewidth}
\includegraphics[width=6cm]{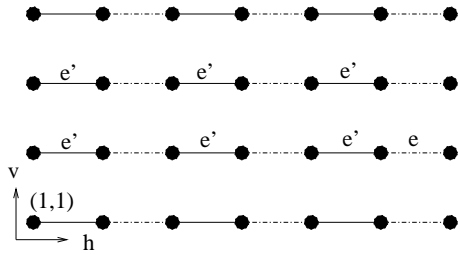}
\caption{\label{fig-sign}For fixed $e$ ($h(L,e)$ is even) in a given
  layer $L$ are marked
  all possible $e'$ ($h(L,e')$ is odd) to be in pair with $e$.}
\end{minipage}
\begin{minipage}{0.61\linewidth}

\begin{definition}
\label{R(Q)}
Let us now define a {\it set of pairs of edges} $R(Q)$.
We let $H$ be the set of all horizontal edges. For each $x\leq m_1$
we let $W_x$ consist of all width edges with their x-coordinate equal to $x$.
We call sets $H$ and $W_x$ {\it layers} of $Q$. It is convenient to depict
the edges of a layer as the horizontal edges of a square grid, where lower
left corner is the smallest in the fixed ordering of vertices. The
first column of the layer $H$ is ordered according to the fixed
ordering of vertices. Size of layer $H$ is $m_1 \times km_2$, sizes of
all $W_x$ are $m_2 \times k$.

If $e$ is an edge
of such a layer $L$ then let $h(L,e)$ ($v(L,e)$ respectively) denote
the horizontal (vertical respectively) coordinate of the first vertex
of $e$ in $L$. For a layer $L$ we let $R(L)$ be the set of all pairs
$(e,e')$ of edges of $L$ such that $h(L,e')< h(L,e), v(L,e)\leq
v(L,e') \leq v(L,e)+1$, and $h(L,e)$ is even while $h(L,e')$ is
odd, example on Fig. \ref{fig-sign}. Finally we let $R(Q)$ to be the
union of $R(L)$ over all layers of $Q$.
\end{definition}

\end{minipage}
\end{figure}

\section{Statement of the main results: The Dimer Problem}
\label{sec.statD}
Our result here is a simplification 
of the main result of \cite{L1}. The reader
is heartily encouraged to check the original statement for a comparison.
Before we write the main theorems let us introduce some necessary notation.

\subsection*{Determinants}
\label{sub.det}
$Q$ is a bipartite graph, which means that its vertices may
be partitioned into two sets $V_1,V_2$ such that if $e$ is an edge
of $Q$ then $|e\cap V_1|=|e\cap V_2|=1$.
We have $|V_1|=|V_2|=m_1m_2k/2$. Let ${\mathcal Z}$ be the square
$(|V_1| \times |V_2|)$ matrix defined by ${\mathcal Z}_{ij}=0$ if
$\{i,j\}$ is not any edge of $Q$. If $e=\{ij\}$ is an edge of $Q$
($i \in V_1$, $j \in V_2$) then ${\mathcal Z}_{ij}=\alpha^{w(ij)}$ if $i<j$ (in
fixed linear order) and  ${\mathcal Z}_{ij}= - \alpha^{w(ij)}$ if $i>j$.

We will consider matrix ${\mathcal Z}$ with its rows and columns ordered
in agreement with the fixed ordering
and we will assume that $V_{111}\in V_1$.
A {\it signing} of a matrix is obtained by multiplying some of
the entries of the matrix by $-1$.

An orientation of a graph $G=(V,E)$ is a {\it digraph} $D=(V,A)$
 obtained from $G$ by assigning an orientation to each edge of $G$,
i.e. by ordering the elements of each edge of $G$. The elements of $A$ are called {\it arcs}.
We say that signing $Z$ of ${\mathcal Z}$ corresponds to orientation $D$
of $Q$ if $Z_{ij}= -{\mathcal Z}_{ij}$ for $(ij)\in D$ and $i>j$.
For arc $(ij)$ of orientation $D$ of $Q$ we let
$\sign(D,(ij))= 1$ if $i<j$, and $\sign(D,(ij))=-1$ otherwise.

\begin{theorem}
\label{thm.uno}
The dimer partition function ${\mathcal P}(Q, \alpha)$ is
$$
{\mathcal P}(Q, \alpha)=2^{-C}\sum_D(-1)^{|\{(e,e')\in R(Q); \hspace{2mm} \sign(D,e)= \sign(D,e')=-1\}|}
{\rm det}(Z(D)),
$$
where the sum is over all orientations of $Q$ with all vertical edges positive,
and $C=km_1(m_2-1)/2 + km_2(m_1-1)/2$.
\end{theorem}

\noindent Knowing the proof Theorem \ref{thm.uno} can be easily
rewritten as

\begin{theorem}
\label{thm.dos}
$$
{\mathcal P}(Q, \alpha)=2^{C}\alpha^{w(M)}- (2^{C}+1)\E,
$$
where $M$ is the unique perfect matching of $Q$ consisting of vertical edges
only and $\E$ equals the average of ${\rm det}(Z(D))$ over all orientations $D$
of $Q$ satisfying: $|\{(e,e')\in R(Q); \hspace{2mm} \sign(D,e)=
\sign(D,e')=-1\}|$
is even (and again all vertical edges are positive).
\end{theorem}

There are $2^{2C}$ orientations to be sum
over in Theorem \ref{thm.uno} and $2^{C-1} (2^C+1)$ terms in Theorem
\ref{thm.dos} to be averaged over. Therefore the exact numerical
analysis is probably not possible. We just tested
Theorem~\ref{thm.uno} for small planar lattices. However it could be
interesting to study statistical
properties of determinants ${\rm det}(Z(D))$. Let us remark here also that
Theorem \ref{thm.uno} for a planar graph (2D Ising model, $m_1=1$) is
still nontrivial, a question remains
how to reduce it to the Kasteleyn's one determinant. And would it be
possible to apply similar reduction even for the non-planar graph?

\section{Statement of the main results: The Ising Problem}
\label{sec.stmI}

The theory described in section \ref{sec.draw} combined with the results of \cite{S, L2}
yields an expression of the 3D Ising partition function as a linear
combination of products over aperiodic closed walks on $Q$. We get a
remarkably simple formula, analogous to the statement for the
dimer partition function. 

\subsection*{Products over aperiodic closed walks}
\label{sub.prod}
Let $G=(V,E)$ be a planar graph embedded in the plane and for each edge $e$ let $x_e$ be an
associate variable. Let
$A=(V,A(G))$ be an arbitrary orientation of $G$.
If $e\in E$ then $a_e$ will denote the orientation of $e$ in $A(G)$
and $a^{-1}$ will be the reversed directed edge to $a$. We let $x_{a_e}= x_{a_e^{-1}}=x_e$.
A circular sequence $p=v_1,a_1,v_2,a_2,...,a_n,(v_{n+1}=v_1)$ is called
{\it non-periodic closed walk} if the following
conditions are satisfied:
$a_i \in \{a_e, a_e^{-1}: e\in E\}$, $a_i\neq a_{i+1}^{-1}$ and $(a_1,...,a_n)\neq Z^m$
for some sequence $Z$ and $m>1$.
We let $X(p)=\prod_{i=1}^n x_{a_i}$.
We further let $\sign(p)=(-1)^{1+n(p)}$,
where $n(p)$ is a {\it rotation number} of $p$,
i.e. the number of integral revolutions of the tangent vector.
Finally let $W(p)=\sign(p)X(p)$.

Now assume $p$ is a closed aperiodic walk in 3D cubic lattice $Q$. Can we define its rotation number?
The following construction provides a solution. 

Assume some vertices of planar graph $G$ embedded in the plane of degree $4$ are marked.
Then we call a walk $p$ {\it correct} if it satisfies the 'crossover condition'
at each marked vertex, i.e. it never enters and exits a marked vertex
in a pair of neighbouring edges along the vertex.

We naturally get a planar graph from $Q$ if we consider $Q$
drawn as bold $Q(3,3,2)$ in Figure~\ref{cube} and put a new vertex to each edge-crossing. These new
vertices will form the set of the marked vertices. Clearly there is 
a natural bijection which associates to each walk $p$ of $Q$ the corresponding correct 
walk $p'$  in the new graph and we define rotation of $p$ to be equal to rotation of $p'$.  
This defines $W(p)$ for each closed aperiodic walk $p$ in $Q$.

There is a natural equivalence on non-periodic closed walks:
$p$ is equivalent with reversed $p$.
Each equivalence class has two elements and will be denoted by $[p]$.
We let $W([p])=W(p)$ and note that this definition is correct since
equivalent walks have the same sign.

We denote by $\prod(1+W([p])$ the formal infinite product of $(1+W([p])$ over
all equivalence classes of non-periodic closed walks of $G$.
 
Let $D$ be an orientation of $Q$. We further let
$$
\prod_D(1+W([p])= \prod(1+W([p])|_{x_e= \sign(D,e)\alpha^{w_e}}.
$$

\begin{theorem}
\label{thm.tres}
Generating function of even subgraphs is
$$
{\mathcal H}(Q,\alpha)= 2^{-C}\sum_D (-1)^{|\{(e,e')\in R(Q); \hspace{2mm} \sign(D,e)= \sign(D,e')=-1\}|}
\prod_D(1+W([p]),
$$
where the sum is over the orientations of $Q$ with all the vertical edges positive.
\end{theorem}

\noindent {\bf Remark.} Theorem \ref{thm.tres} expresses the 3D Ising partition function as a 
linear combination of infinite products over aperiodic closed walks. For planar lattice this reduces 
to just one infinite product (conjectured by Feynman, proved in \cite{S}) which provides 
an equivalence of planar Ising model with the quantum field theory of free fermion. There 
have been several attempts to generalize this equivalence for 3D. Earlier papers 
(e.g. \cite{D}) attempt to replace closed aperiodic walks by 2D surfaces. Recent 
promising development mentioned in the introduction deals with discrete Riemann surfaces 
and the theta function:

It is known that the planar Ising partition function in the thermodynamical limit
behaves like a 'common term' times a Riemann theta function.
For torus, \cite{AMV} asserts that in the thermodynamic
limit and near to criticality, 
the Ising partition function behaves like a 'common term' times a linear
combination of four Riemann theta functions corresponding to torus.
 In \cite{M} and \cite{CM1, CM2}, there is evidence that  
for the critical Ising model, the dependence of the determinants of
adjacency matrices on the Kasteleyn orientations is exactly the same as
the dependence of the determinants of the Dirac operator, of the corresponding conformal
field theory, on the spin structures of the 2D Riemann surface; this is given
in terms of theta functions of half-integer characteristics.

We believe that theorem \ref{thm.tres} may provide a new insight into these efforts.

\section{Drawing of cubic lattices}
\label{sec.draw}

In this section we describe how we draw cubic lattices on 2-dimensional surfaces.

\subsection{Theory of generalised g-graphs.}
\label{sub.ge}

\begin{definition}
\label{def.surface}
A {\it surface polygonal representation} $S_g$ of an orientable 2D surface of genus $g$
consists of a {\it base} $B_0$ and $2g$
{\it bridges} $B^i_j$, $i=1,...,g$ and $j=1,2$, where

\begin{itemize}
\item [i)]  $B_0$ is a convex $4g$-gon with vertices
$a_1, ...,a_{4g}$ numbered clockwise;

\item [ii)]  $B^i_1$, $i=1,\dots,g$, is a $4$-gon
with vertices $x^i_1, x^i_2, x^i_3, x^i_4$ numbered clockwise. It is glued
with $B_0$ so that the edge $[x^i_1,x^i_2]$ of $B^i_1$ is identified with
the edge $[a_{4(i-1)+1},a_{4(i-1)+2}]$ of $B_0$ and the edge $[x^i_3,x^i_4]$ of
$B^i_1$ is identified with the edge $[a_{4(i-1)+3},a_{4(i-1)+4}]$ of $B_0$;

\item [iii)] $B^i_2$, $i=1,\dots,g$, is a $4$-gon with vertices
   $y^i_1, y^i_2, y^i_3, y^i_4$ numbered clockwise. It is glued
 with $B_0$ so that the edge $[y^i_1,y^i_2]$ of $B^i_2$ is identified with
 the edge $[a_{4(i-1)+2},a_{4(i-1)+3}]$ of $B_0$ and the edge
 $[y^i_3,y^i_4]$ of $B^i_2$ is identified with
 the edge $[a_{4(i-1)+4},a_{4(i-1)+5}]$ of $B_0$. Indexing is $ {\rm
   mod}\,  4g $.
\end{itemize}
\end{definition}

\begin{definition}
\label{def.embed}
A graph $G$ is called a {\it $g$-graph} if it is embedded on $S_g$
 so that all the vertices belong to the base $B_0$, and
 the embedding of each edge uses at most one bridge.
We denote the set of the edges embedded entirely on the base by
 $E_0$ and the set of the edges embedded on each bridge $B^i_j$ by $E^i_j$,
$i=1,\dots,g,$  $j=1,2$.
 \noindent
Moreover the following conditions need to be satisfied.
\begin{enumerate}
\item[1.]
 the outer face of $G_0=(V,E_0)$ is a cycle,
 and it is embedded on the boundary of $B_0$,
\item[2.]
if $e\in E^i_1$ then $e$ is embedded entirely on $B^i_1$ with
one end-vertex belonging to $[x^i_1,x^i_2]$ and the other one
 to $[x^i_3,x^i_4]$. Analogously for $e\in E^i_2$.
\end{enumerate}
\end{definition}

We need a generalization of the notion of a $g$-graph.

\begin{definition}
\label{def.geng}
Any graph $G$ obtained by the following construction will be called
{\it generalized g-graph}.
\begin{enumerate}
\item[1.] Let $g=g_1+...+g_n$ be a partition of $g$ into positive integers.
\item[2.] Let $S_{g_i}$ be a polygonal representation of a surface  of genus
          $g_i$, $i=1,...,n$. Let us denote the basis and the bridges of $S_{g_i}$ by $B_0^i$ and
          $B^i_{j,k}$, $i=1,...,n$, $j=1,...,g_i$ and $k=1,2$.
\item[3.] For $i=1,...,n$ let $H_i$ be a $g_i$-graph with the property
          that the subgraph of $H_i$ embedded on $B^i_0$ is a cycle, embedded on the boundary
          of $B^i_0$. Let us denote it by $C^i$.
\item[4.] Let $G_0$ be a $2$-connected graph properly embedded on the plane and let $F_1,...,F_n$
          be a subset of the faces of $G_0$. Let $K^i$ be the
          cycle bounding $F_i$, $i=1,...,n$. Let each $K^i$ be isomorphic
          to $C^i$.
\item[5.] Then $G$ is obtained by gluing the $H_i$'s into $G_0$ so
          that each $K^i$ is identified with $C_i$.
\end{enumerate}
\end{definition}

\noindent {\bf Orientations.}
Let $G$ be a  $g$-graph and let $G^i_j=(V,E_0\cup E^i_j)$.
An orientation $D_0$ of $G_0$ such that each inner face of
each $2$-connected component of $G_0$ is clockwise odd in $D_0$
is called a {\it basic orientation} of $G_0$.
Note that a basic orientation always exists for a planar graph.
Further we define the orientation $D^i_j$ of each $G^i_j$ as follows:
We consider $G^i_j$ embedded on the plane by the planar projection of $E^i_j$
outside $B_0$, and complete the basic orientation $D_0$ of $G_0$ to an orientation
 of $G^i_j$ so that each inner face of each $2$-connected
 component of $G^i_j$ is clockwise odd.
The orientation $-D^i_j$ is defined by complete reversing of the orientation $D^i_j$ of
$G^i_j$.

Observe that after fixing a basic orientation $D_0$,
the orientation $D^i_j$ is uniquely determined for each $i,j$.

\begin{definition}
\label{def.r}
Let $G$ be a  $g$-graph, $g \geq 1$. An orientation $D$ of $G$
which equals the {\it basic orientation} $D_0$ on $G_0$ and which equals
$D^i_j$ or $-D^i_j$ on $E^i_j$
 is called {\it relevant}. We define its {\it type}
 $r(D) \in \{+1,-1\}^{2g}$ as follows:
For $i =0,\dots,g-1$ and $j =1,2$, $r(D)_{2i + j}$ equals  $+1$ or $-1$
according to the sign of $D^{i+1}_j$ in $D$.

Moreover we let $c(r(D))$ equal  the product
of $c_i$, $i=0,...,g-1$, where $c_i=c(r_{2i+1},r_{2i+2})$ and
$c(1,1)=c(1,-1)=c(-1,1)=1/2$ and $c(-1,-1)=-1/2$.
\end{definition}

Observe that $c(r(D))=(-1)^n 2^{-g}$, where
$n=|\{i;r_{2i+1}=r_{2i+2}=-1\}|$.

For each generalized g-graph $G$ we can define $4^g$ relevant orientations
$D_1,...,D_{4^g}$ with respect to a fixed basic orientation of $G_0$,
 and coefficients $c(r(D_i))$, $i=1,...,n$
in the same way as for a g-graph. Now we write theorem proved in \cite{GL}
which will be essential in proof of Theorem~\ref{thm.uno}.

\begin{theorem}
\label{thm_GL}
   Let G be a generalized g-graph with a perfect matching $M_0$ of $G_0$. Let
   $D_0$ be a basic orientation of $G_0$. If we order the vertexes of $G$ so
   that $s(D_0,M_0)$ is positive then
   $$
              {\mathcal P}(G, \alpha) = \sum_{i=1}^{4^g} c(r(D_i))
              Pf_{G}(D_i,\alpha),
   $$
   where $D_i$ are all the relevant orientations of $G$.
\end{theorem}

Definition of $s(D_0,M_0)$ follows from definition of {\it Pfaffian}. 

Let $G=(V,E)$ be a graph with $2n$ vertices and $D$ an orientation of $G$. 
Denote by $A(D)$
the skew-symmetric matrix with the rows and the columns indexed by 
$V$, where $a_{uv}=\alpha^{w({u,v})}$ in case $(u,v)$ is an arc of $D$, 
$a_{u,v}= -\alpha^{w({u,v})}$ in case $(v,u)$ is an arc of $D$, and $a_{u,v}=0$
otherwise. 

\begin{definition}
\label{def_pf}
The {\it Pfaffian} is defined as 
$$
    Pf_G(D,\alpha)=\sum_{P} s^*(P)a_{i_1j_1} \cdots a_{i_nj_n},
$$
where $P=\{\{i_1j_1\},\cdots,\{i_nj_n\}\}$ is a partition of the 
set $\{1,\dots,2n\}$ into pairs, $i_k<j_k$ for $k=1,\dots,n$, and 
$s^*(P)$ equals the sign 
of the permutation $i_1j_1\dots i_nj_n$ of $12\dots (2n)$.

Each nonzero term of the expansion of the Pfaffian
equals $\alpha^{w(P)}$ or $-\alpha^{w(P)}$ where $P$ is a perfect
matching of $G$. If $s(D,P)$ denote the sign of the term
$\alpha^{w(P)}$ in the expansion, we may write
$$
    Pf_G(D,\alpha)=\sum_{P} s(D,P)\alpha^{w(P)}. 
$$
\end{definition}

\subsection{Cubic lattices as generalized g-graphs.} 

In this subsection we will describe how to realize $Q(m_1,m_2,k)$ as
a subgraph of a generalized g-graph. Let us denote by $Q^L$ the larger
lattice $Q(m_1, m_2, 2k+1)$.
\\

\begin{figure}[ht]
 \includegraphics[width=9cm]{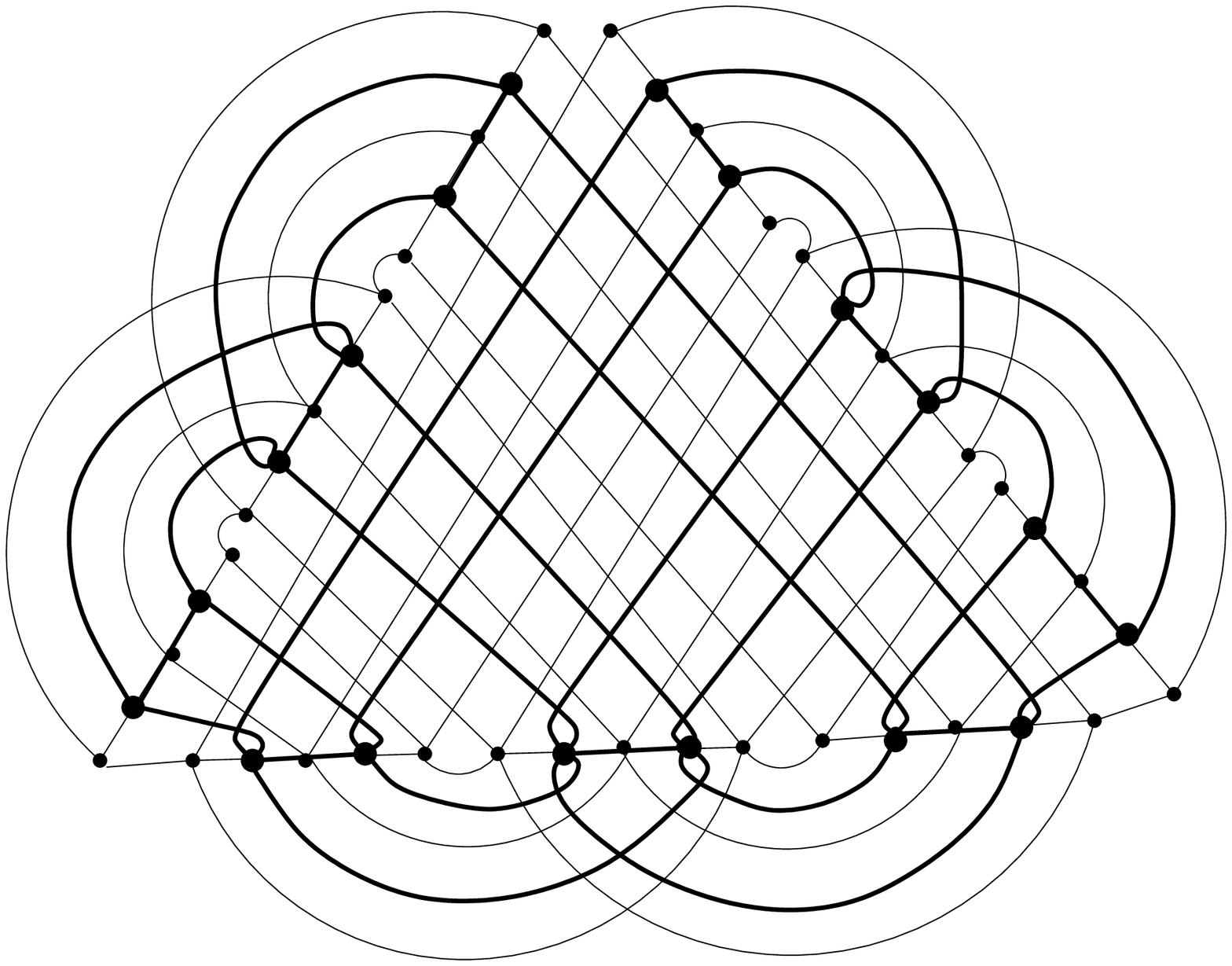}
\caption{\label{cube} Drawing of a cube $Q^L=Q(3,3,3)$ in a plane.}
\end{figure}

\noindent {\bf How to draw $Q^L$ on the plane.}
First draw the paths $V_{xy}$ along a cycle in the linear order as in
Def. \ref{lin_order}.
Next, draw the horizontal edges inside this cycle, and the width edges
outside of this cycle as depicted in Fig.~\ref{cube} below where $Q^L=Q(3,3,5)$
is properly drawn.
The figure also shows how $Q = Q(3,3,2)$ is embeded in 
$Q^L=(3,3,5)$.

We keep the following rule: the interiors of the curves
representing $h_{xyz}$ and $h_{(x+1)yz}$ ($w_{xyz}$ and $w_{x(y+1)z}$
respectively)
intersect if and only if $z$ is even. We denote by $C(e)$ the curve
representing edge $e$.

\noindent {\bf Now we  modify $Q^L$ into a generalized g-graph $Q'$.}
We introduce new vertices to some edge-crossings of $Q^L$ and delete
and subdivide some edges.
In this way we obtain
a generalised g-graph $Q'$ with an even subdivision of $Q(m_1, m_2, k)$ as its subgraph.
The construction is analogous
and independent for each plane, and so we describe it only for a plane $W_x$.

\begin{figure}[ht]
 \includegraphics[width=11cm]{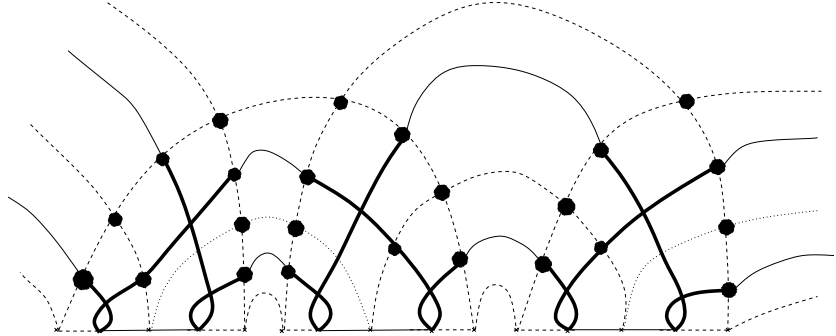}
\caption{\label{cons_Q}  Construction of $Q'$ for part of the plane $W_x$
  ($x$ odd), $V_{x(y-1)}, V_{xy}, V_{x(y+1)}$ ($y$ even).}
\end{figure}

The construction is described by Fig.~\ref{cons_Q} for the edges between $V_{x(y-1)}, V_{xy}$ and
between $V_{xy}, V_{x(y+1)}$, for $x$ odd and $y<m_2-1$ even.

\begin{enumerate}
 \item For each $y$ even let $Aux_1=\{w_{xyz}; z$ odd $\}$.
 For each edge $e$ of $Aux_1$ introduce a new vertex to each
intersection of $C(e)$ with the curves representing the edges
of $W_{x(y-1)} \cup W$, where $W=W_{x(y+1)}$ in case $y < m-1$ and
$W=\emptyset$ otherwise. By this operation, each  $e \in Aux_1$
is replaced by a path. Call each edge of this path {\it auxiliary}.

\item For each $y$ even let $Aux_2=\{w_{x(y-1)1}, w_{x(y-1)n}\}\cup A$, where
$A=\{w_{x,(y+1)1}, w_{x(y+1)n}\}$ in case $y < m-1$ and $A=\emptyset$ otherwise.
For each edge $e$ of $Aux_2$ introduce a new vertex to each
intersection of $C(e)$ with the curves representing the edges
of $W_{xy}$. Hence each  $e \in Aux_2$
is replaced by a path. Call each edge of this path {\it auxiliary}.
 For each $y$ even the edges $v_{xy1}, v_{xy(n-1)}$
and also $v_{x(y+1)1}, v_{x(y+1)(n-1)}$ will also  be called {\it auxiliary}.
 In Figure~\ref{cons_Q}, the auxiliary edges are represented by  dashed lines.

\item  The edges $w_{xyz}$, $y$ even and $z$ even will be called
{\it relevant} for $Q$. If $y < m-1$ then the relevant edges are subdivided
by two vertices (added in 2.) into three edges of $Q'$. The middle one will
be called {\it special} and the other two {\it long}.
 If $y=m-1$ then the relevant edge $w_{xyz}$ is subdivided by one vertex
into two edges of $Q'$. The one incident to $V_{xm}$
will be called {\it special} and the other one {\it long}.
 If $e$ is a relevant edge of $Q$, then we choose a
initial long edge $f$ and we let $w(e)=w(f)$. We let the weight of the
special edge and of the remaining long edge be equal to 0.

\item  The edges of $W_{x(y-1)}\cup W$ also got subdivided by new
vertices introduced in step~1 and step~2.

\item  We delete all edges of the paths
obtained from $w_{x(y-1)z}$ and $w_{x(y+1)z}$,
$1 < z < n$ odd, as well as we delete all vertexes in intersections of those
edges with auxiliary edges.
In Figure~\ref{cons_Q}, the deleted edges are represented by dotted lines.

\item  Each edge $e \in \{w_{x(y-1)z}, w_{x(y+1)z}; z$ even $\}$,
is subdivided by new vertices introduced in step 1 into a path.
We let the weights assigned to the edges of the path equal 0 except of one
initial edge whose weight is let equal $w(e)$.
The edge $e$ of this path such that the interior of $C(e)$ does not intersect interior of
any curve representing a long edge will also be {\it special}.
The others are called {\it short}.

\item  All vertical edges which are not auxiliary (see Fig.\ref{cons_Q})
will be called {\it special}.
In Figure~\ref{cons_Q}, the special edges are represented by normal lines.

\item 
Let $Aux$ denote the set of all auxiliary edges. Then deletion of $Aux$ results
in a subdivision of $Q_{m_1m_2k}$. We subdivide some special edges (vertical and border
width ones) so that the graph
$\ti Q = Q'-Aux$ is an even subdivision of $Q_{m_1m_2k}$. All these new edges will be
special, and we set their weights equal~0.
\end{enumerate}

This finishes the construction. In Figure~\ref{cons_Q}, the long and short edges
are represented by fat lines.

{\bf $Q'$ is a generalised g-graph.} Its planar part $Q^p$ is set of all the
auxiliary and special edges. Other edges (i.e. the short and long) are drawn on
a face of $Q^p$ and they may be drawn onto a pair of bridges above this face.
One bridge contains one long edge, and the other bridge contains all the short
edges.

Now let us set appropriate weights to all the auxiliary edges. In order to do
that first subdivide and add some auxiliary edges and vertices to $Q^p$ 
in such a way that there exists a matching $M_{Aux}$ of
$V(Q')-V(\ti Q)$ consisting of auxiliary edges, and a perfect matching $M_0$ of $Q^p$. 
 We set weights of edges from
$M_{Aux}$ as $w(e)=0$. Weights of other auxiliary edges $w(e)= -\infty$.
With such properties of $Q'$ we have an important result: ${\mathcal
  P}(Q',\alpha)= {\mathcal P}(Q(m_1,m_2,k),\alpha)$.

Now we introduce a {\it basic orientation} $D^p$ of $Q^p$ in order to prepare
our stage to use the Theorem~\ref{thm_GL} for the generalised g-graph $Q'$.
Orientation $D^p$ has the following properties:
\begin{enumerate}
\item[1.]
$D^p$ on special edges is in agreement with the natural ordering
(Def. \ref{lin_order}).
\item[2.] All the signs $s(D^p,P)$ in Pfaffian $Pf_{Q^p}(D^p,\alpha)$
(Def.~\ref{def_pf}) are positive.
\item[3.]
The orientation of edges on a bridge has positive sign if and only if
it is in agreement with the natural ordering (Def.~\ref{lin_order}).
\end{enumerate}
The construction of such a $D^p$ is possible according to Kasteleyn
\cite{K}.

\section{The proofs}


\subsection{\bf Proof of Theorem \ref{thm.uno}:} 

\noindent We  use
Theorem \ref{thm_GL} for the graph $Q'$. 
Then we may write
\begin{eqnarray}
   {\mathcal P}(Q{(m_1m_2k)},\alpha) &=& {\mathcal P}(Q',\alpha) =
   \sum_{i=1}^{4^g} c(r(D_i)) Pf_{Q'}(D_i,\alpha) = \nonumber \\   
   &=& \sum_{i=1}^{4^g}
   c(r(D_i)) Pf_{\ti Q \cup M_{Aux}}(D_i,\alpha) =  \sum_{i=1}^{4^g}
   c(r(D_i)) Pf_{Q}(D^*_i,\alpha), \nonumber   
\end{eqnarray}
where the first equality is the result of previous section. The second
equality holds because of the Theorem \ref{thm_GL}. The third equality
holds because of the setting of the  weights of edges in $Q'$ in the
previous section. And finally, orientation $D^*_i$ being induced on
$Q$ by $D_i$ on $\ti Q$, the fourth equality follows from the
three facts: setting of the weights of edges, $\ti Q$ is even
subdivision of $Q$, it is possible to permute vertexes covered by
$M_{Aux}$ so that the sign is correct.

In Section \ref{sub.det} we have defined for a bipartite graph $G$ a matrix
$Z(D)$ depending on a orientation $D$. It is easy to check that for
$Q$ holds $Pf_{Q}(D^*_i,\alpha)={\rm det} (Z(D^*_i))$.

To proceed in proof of Theorem \ref{thm.uno} we have to establish the
dependence of $c(r(D_i))$ on the orientation $D^*_i$ of $Q$ and
specify the orientations we sum over in the language of $Q$ not $Q'$.

Each relevant orientation $D_i$ of $Q'$ is determined by the fixed basic 
orientation $D^p$ of $Q^p$, and by a pair of signs for each pair of bridges.
Each pair of bridges is associated with one long edge and set of short
edges of $Q'$. Hence these signs
may be given by specifying $(d_{D_i}^1(e),d_{D_i}^2(e)) \in \{+-\}^2$, 
for each long edge $e$,
where $d_{D_i}^1(e)$ denotes the sign of the bridge containing $e$, and
$d_{D_i}^2(e)$ denotes the sign of the other bridge containing the set
of short edges.

The relevant edges $w_{x(m_2-1)z}(Q{(m_1m_2k)})$ 
and $h_{(m_1-1)yz}(Q{(m_1m_2k)})$ are associated with only
one long edge of $Q'$. If $e$ is such relevant edge of $Q$,
we will call it {\it border edge}, and we denote by $e_1$ the corresponding
long edge. We let $d_{D_i}(e)=(d_{D_i}^1(e_1),d_{D_i}^2(e_1),+,+)$.
Each relevant non-border edge $e$ of $Q$ has two long edges
$e_1,e_2$ associated with it. We let
$d_{D_i}(e)=(d_{D_i}^1(e_1),d_{D_i}^2(e_1),d_{D_i}^1(e_2),d_{D_i}^2(e_2)))$.
A {\it relevant vector} is any element $r$ of $[\{+,-\}^4]^{\mathcal R}$
such that $r(e)_3=r(e)_4= +$ for each relevant border edge $e$ of $Q$.

From the definition of $c(r(D_i))$ and from properties of the
generalized g-graph $Q'$ of genus $g$ follows that $c(r(D_i))=2^{-g} (-1)^n$,
where $n$ is number of faces of $Q^p$ above them edges on both bridges
are oriented in negative sense, i.e. against fixed ordering. 
There is a natural bijection between  relevant orientations of $Q'$
and relevant vectors. 
If $r$ is a relevant vector, then let $D_i(r)$ denote the corresponding 
relevant orientation of $Q'$ and let $\sgn(r)$ of a relevant vector $r$ 
be calculated as follows: 
$\sgn(r)=2^g c(r(D_i)) =(-1)^{|\{(e,i); \quad i=0,1; \quad
r(e)_{2i+1}=r(e)_{2i+2}=-1\}|}$.

There is $C_r = km_1(m_2-1)/2 +km_2 (m_1 -1)/2$ relevant edges and
$C_b=k(m_1+m_2)$ border edges. Genus of $Q'$ is therefore
$g=2C_r-C_b$. There are $4^{2C_r-C_b}$ relevant vectors.

Let us consider a set $\mathcal D$ of relevant orientations $D_i$ such that
there is at least one relevant non-border edge $e$ with $r(e)_2 \neq r(e)_4$,
i.e. orientation of bridges with short edges incident to $e$ are
different.  

\begin{proposition}
  \label{sum_zero} Sum of contribution of all orientations from $\mathcal D$
  to ${\mathcal P}(Q,\alpha)$ is zero.
\end{proposition}
\noindent {\bf Proof:} 
For given orientation $D \in \mathcal D$ define
a set $A_D = \{ e \quad {\rm relevant}, r(e)_2 \neq r(e)_4 \}$. Now
let fix a set o relevant edges $A_f$ and consider all orientations
$D_f$ such that $A_{D_f} = A_f$. Fix also one relevant edge $e_f\in
A_f$. 

Remark: $c(r(D_1)) +c(r(D_2)) =0$ if $r(D_1)$ and $r(D_2)$ differs
only in one relevant edge $e$ in such a way that $r_{D_1}(e)_2
= r_{D_2}(e)_2 \neq r_{D_1}(e)_4=  r_{D_2}(e)_4$ and  $r_{D_1}(e)_1= -
r_{D_2}(e)_1$, $r_{D_1}(e)_3 = - r_{D_2}(e)_3$. It is also that
induced orientations $D^*_1 = D^*_2$.

Using this remark for $e=e_f$ we have for every $A_f$: 
$$
      \sum_{D_i \in D_f} c(r(D_i)) Pf_{Q}(D^*_i,\alpha) = 0.               
$$
The claim in Proposition \ref{sum_zero} follows directly. 
\done 
\medskip

Let us call all $D_i \not\in \mathcal D$ {\it useful orientations} and
from now we sum only over such orientations. There is $ 2^{3C_r-2C_b}$
of useful orientations. Similarly we define useful relevant vectors.

\begin{proposition}
  \label{all_D} Each orientation $D$ of $Q$, which agrees
  with the fixed order on vertical edges, corresponds to $2^{C_r-C_b}$ of
  useful orientations $D_i$ of $Q'$ and all those $D_i$ have the same
  sign $c(r(D_i))$.
\end{proposition}
\noindent {\bf Proof:} Let $r$ be a useful vector. Then $D(r)$
  determines uniquely $r(e)_2$ and $r(e)_4$ for each relevant edge $e$
  and also $r(f)_1$ for each relevant border edge $f$. Hence $D(r)$
  determines uniquely $r(f)$ for each relevant border edge
  $f$. Moreover $D(r)$ determines uniquely the product $r(e)_1\times
  r(e)_3$ for each relevant non-border edge $e$. Since there are
  $C_r-C_b$ relevant non-border edges, there is $2^{C_r-C_b}$
  orientations $D_i$ corresponding to it.

  Let $r,s$ be useful and both lead to the same $D$ of $Q$. Then
  $r(e)_2=s(e)_2=s(e)_4=r(e)_4$ for each relevant non-border edge $e$
  and $r(e)_2=s(e)_2$ and $s(e)_1=r(e)_1$ for each relevant border
  edge. This implies that $\sgn(r)=\sgn(s)$.
\done 
\medskip

Concluding from what we did till now we can write the dimer partition
function as
$$
  {\mathcal P}(Q,\alpha) = 2^{-C_r} \sum_{D} \sgn(D) {\rm det}(Z(D)),     
$$
where the sum is over all orientations of $Q$ with all vertical edges
positive. At this point the only thing missing in proof of Theorem
\ref{thm.uno} is to concretize $\sgn(D)$.
\begin{proposition}
  \label{sgn_D} $\sgn(D)=(-1)^{|\{(e,e')\in R(Q); \hspace{2mm}
    \sign(D,e)= \sign(D,e')=-1\}|}$, where the set of edges $R(Q)$ is
    defined in \ref{R(Q)}.
\end{proposition}
\noindent {\bf Proof:} Let us remind that $\sgn(D)=(-1)^n$ where $n$
    is number of faces with both bridges above them negative. Each
    non-border relevant edge $e$ corresponds to two faces with
    bridges. The contribution of the two corresponding faces to sign
    could be negative only if orientation of $e$ is negative
    ($r(e)_1\neq  r(e)_3 $). And moreover if orientation of
    corresponding shot edges is negative ($r(e)_2= r(e)_4 = -1 $).
    It follows from the construction that this happen only if product
    of orientation of all $e'$ (as defined in \ref{R(Q)} ) is
    negative.
    Considering all the relevant edges we get the claim of Proposition
    \ref{sgn_D} 
\done \medskip

\subsection{\bf Proof of Theorem \ref{thm.dos}:} 

\begin{proposition}
   \label{av}
   The average of ${\rm det}(Z(D))$ over all orientations of $Q$ with all
   vertical edges fixed positive equals $\alpha^{w(M)}$, where $M$ is
   perfect matching of $Q$ consisting only from vertical edges.
\end{proposition}
\noindent {\bf Proof:}
By the linearity of expectation 
the contribution of other than vertical edges cancel out when we calculate
the average of ${\rm det}(Z(D))$. Since $Q{(m_1m_2k)}$ has exactly one
perfect matching consisting of vertical edges only, Proposition
\ref{av} follows.
\done \medskip

There is $2^{2C}$ orientation D in \ref{thm.uno} to be summed over.
And $2^{C-1}(2^C+1)$ of them have positive sign. Proof: Looking at
Figure \ref{fig-sign} we can see that as soon as there is at least one
relevant (shaded) edge negative there is so many orientation with
positive as with negative sign. Only if all the relevant edges are
positive the sign is also positive. It means there is $2^C$ more
positive orientation then negative. 

By Theorem \ref{thm.uno} and Proposition \ref{av} we have then 
$$
  {\mathcal P}(Q,\alpha)=2^{-C}[-2^{2C} \alpha^{w(M)} +2\mathcal
  E (2^{C-1}(2^{C}+1))]= -2^{C} \alpha^{w(M)} +  \mathcal
  E (2^{C}+1),
$$
where $\mathcal E$ equals the average of ${\rm det}(Z(D))$ over all $D$
of $Q$ with positive sign. And Theorem \ref{thm.dos} is proved.
\done \medskip

\subsection{\bf Proof of Theorem \ref{thm.tres}:}

We use a straightforward reformulation of a theorem of \cite{L2}:

\begin{theorem}
\label{thm.prd}
  Let G be a generalized g-graph. Then
   $$
              {\mathcal H}(G, \alpha) = \sum_{i=1}^{4^g} c(r(D_i))
             \prod_{D_i}(1+W([p]) ,
   $$
   where $D_i$ are all the relevant orientations of $G$, and rotation of $p$
and $\prod_{D_i}(1+W([p])$ are defined in the same way as described for $Q$
in section \ref{sub.prod}.
\end{theorem}

\noindent Proof of this theorem could be found in \cite{L2} and for planar graphs in \cite{S}.

Let $D_i, i=1,\dots, 4^g$ be the relevant orientations of generalised g-graph $Q'$.
If we apply theorem \ref{thm.prd} to $Q'$, from the choice of weights we get
$$
              {\mathcal H}(Q, \alpha) = \sum_{i=1}^{4^g} c(r(D_i))
             \prod_{D^*_i}(1+W([p]).
$$
The same analysis as in the previous section finishes the proof of Theorem \ref{thm.tres}.

\section{Conclusions}
\label{sec.con}

We express the finite 3D Dimer partition function as a linear combination of determinants 
of oriented adjacency matrices in Theorem \ref{thm.uno}, and the finite 3D Ising partition 
function as a linear combination of products over aperiodic closed walks in Theorem \ref{thm.tres}.

The methodology we use is embedding of cubic lattice on 2D surfaces of large genus. By this embedding 
cubic lattice is transformed to a generalized g-graph. Then we use theorems from \cite{GL, L2}, which 
generalize approach of \cite{K}. 

We believe this approach may relate to recent developments in conformal field theory 
\cite{M,CM1,CM2} and topological string theory \cite{ORV}, and ultimately teach us something 
about 3D basic statistical physics problems.

\medskip
\ifx\undefined\bysame
        \newcommand{\bysame}{\leavevmode\hbox
to3em{\hrulefill}\,}
\fi

\end{document}